\documentclass[usegraphicx]{mn2e}

\input{epsf}
\begin{document}
\pagestyle{headings}
\newcommand{\etal}{{\rm et al.}~}
\newcommand{\hmpc}{$h^{-1}\rm Mpc$}
\newcommand{\kms}{$\hbox{km}\cdot \hbox{s}^{-1}$}
\newcommand{\be}{\begin{equation}}
\newcommand{\ee}{\end{equation}}
\newcommand{\de}{\delta}
\newcommand{\s}{\sigma}
\newcommand{\te}{\theta}
\newcommand{\f}{\frac}
\newcommand{\bfx}{{\mathbf x}}
\newcommand{\bfr}{{\mathbf r}}
\newcommand{\bfs}{{\mathbf s}}
\newcommand{\bft}{{\mathbf t}}
\newcommand{\bfz}{{\mathbf z}}
\newcommand{\bfy}{{\mathbf y}}
\newcommand{\bfk}{{\mathbf k}}
\newcommand{\bfv}{{\mathbf v}}
\newcommand{\bfq}{{\mathbf q}}
\newcommand{\bfg}{{\mathbf g}}
\newcommand{\bfp}{{\mathbf p}}
\newcommand{\bfu}{{\mathbf u}}
\newcommand{\calD}{{\mathcal D}}
\newcommand{\calF}{{\mathcal F}}
\newcommand{\calO}{{\mathcal O}}
\newcommand{\calQ}{{\mathcal Q}}
\newcommand{\calC}{{\mathcal C}}
\newcommand{\calI}{{\mathcal I}}
\newcommand{\calL}{{\mathcal L}}
\newcommand{\calK}{{\mathcal K}}
\newcommand{\calN}{{\mathcal N}}
\newcommand{\calS}{{\mathcal S}}
\newcommand{\calH}{{\mathcal H}}
\newcommand{\calP}{{\mathcal P}}
\newcommand{\lan}{\langle}
\newcommand{\ran}{\rangle}

\title[Cosmic Velocity Fields]{
Gaussianity of Cosmic Velocity Fields and \\
Linearity of the Velocity--Gravity Relation}
\author[Cieciel\c{a}g et al.]{Pawe\l\ Cieciel\c{a}g,$^1$\thanks{E-mail: 
			      	pci@camk.edu.pl} Micha\l\
			      J. Chodorowski,$^1$\ Marcin Kiraga,$^2$\
			      \newauthor Michael A. Strauss,$^3$\ 
			      Andrzej Kudlicki$\,^1$\thanks{Currently
			      at Univ.\ of Texas Southwestern Medical
			      Center, Dallas, TX 75390-9038 USA}\   
			      and Fran\c cois R. Bouchet$\,^4$
	\\
        $^1$Copernicus Astronomical Center, 
              Bartycka~18, 00-716 Warsaw, Poland \\
        $^2$Warsaw University Observatory, 
		Aleje Ujazdowskie 4, 00-478 Warsaw, Poland\\
        $^3$Princeton University Observatory, 
                      Princeton, NJ 08544-1001  USA \\
	$^4$Institut d'Astrophysique, 98 bis Boulevard Arago, 
			75014 Paris, France \\
        }

\maketitle

\begin{abstract}
We present a numerical study of the relation between the cosmic
peculiar velocity field and the gravitational acceleration field. We
show that on mildly non-linear scales (4--10 \hmpc\ Gaussian
smoothing), the distribution of the Cartesian coordinates of each of
these fields is well approximated by a Gaussian. In particular, their
kurtoses and negentropies are small compared to those of the velocity
divergence and density fields. We find that at these scales the
relation between the velocity and gravity field follows linear theory
to good accuracy. Specifically, the systematic errors in
velocity--velocity comparisons due to assuming the linear model do not
exceed 6\% in $\beta$. To correct for them, we test various nonlinear
estimators of velocity from density. We show that a slight
modification of the $\alpha$-formula proposed by Kudlicki \etal yields
an estimator which is essentially unbiased and has a small variance.

\end{abstract}

\section{Introduction}
\label{sec:intro}

According to the gravitational instability paradigm, structures in the
universe formed by the growth of small inhomogeneities present in the
early Universe.  Gravitational instability gives rise to a coupling
between the density and peculiar velocity fields on scales larger than
the size of clusters of galaxies, the largest bound objects in the
Universe.  On very large, linear scales, the relation between the
density contrast $\de$ and the peculiar velocity $\bfv$ in co-moving
coordinates can be expressed in differential form,
\be
\de(\bfr) = - (H_0 f)^{-1} \nabla \cdot \bfv(\bfr) 
\,,
\label{eq:lindif}
\ee
or in integral form,
\be
\bfv(\bfr)= \bfg(\bfr) \,.
\label{eq:linint}
\ee
Here,
\be
 \bfg(\bfr) \equiv H_0 f \int \frac{{\rm d}^3\bfr'}{4 \pi} 
   \frac{\delta(\bfr')\, (\bfr'-\bfr)} {|\bfr'-\bfr|^3}
 \label{eq:accel}
\ee
is a quantity {\em proportional} to the gravitational
field,\footnote{The linear relation between the peculiar velocity and
the gravity, $\tilde\bfg$, is (e.g., Peebles 1980) $\bfv =
2f/(3H\Omega_m)\, \tilde\bfg$. For simplicity, here we define the {\em
scaled} gravity, $\bfg \equiv 2f/(3H\Omega_m)\, \tilde\bfg$. Then
eq.~(\ref{eq:linint}) follows and $\bfg$ is given by
eq.~(\ref{eq:accel}).} expressed in units of
km$\:\cdot\:$s$^{-1}$. The coupling constant, $f$, carries information
about the underlying cosmological model, and is related to the
cosmological matter density parameter, $\Omega_m$, and cosmological
constant, $\Omega_\Lambda$, by
\be
f(\Omega_m,\Omega_\Lambda) \simeq \Omega_m^{0.6} +
\f{\Omega_\Lambda}{70} \left(1 + \frac{\Omega_m}{2}\right)
\label{eq:f_factor}
\ee
(Lahav \etal 1991).
Hence, comparing the observed density and velocity fields of galaxies
allows one to constrain $\Omega_m$, or the degenerate combination $\beta
\equiv \Omega_m^{0.6}/b$ in the presence of galaxy biasing (e.g.,
Strauss \& Willick 1995 for a review). This comparison is done by extracting the
density field from full-sky redshift surveys (such as the PSCz;
Saunders \etal 2000), and comparing it to the observed velocity field
from peculiar velocity surveys.  The methods for doing this fall into
two broad categories.  One can use equation~(\ref{eq:linint}) to
calculate the predicted velocity field from a redshift survey, and
compare the result with the measured peculiar velocity field; this is
referred to as a velocity-velocity comparison.  Alternatively, one can
use the differential form, equation~(\ref{eq:lindif}), and calculate
the divergence of the observed velocity field to compare directly with
the density field from a redshift survey; this is called a
density-density comparison.  Nonlinear extensions of
equation~(\ref{eq:lindif}) have been developed by a number of workers
(Nusser \etal 1991, Bernardeau 1992, Gramann 1993, Mancinelli \etal
1994, Mancinelli \& Yahil 1995, Chodorowski 1997, Chodorowski \&
{\L}okas 1997, Chodorowski \etal 1998, Bernardeau \etal 1999, Dekel
\etal 1999, Kudlicki \etal 2000, hereafter KCPR; see also the
discussion below). However, very little work has been done to test on
what scales the integral relation, equation~(\ref{eq:linint}) holds,
and how it might be extended into the mildly nonlinear regime; thus
the motivation for this paper.  Attempts have been made to carry out
velocity--velocity comparisons with very large smoothing lengths
(e.g., the inverse Tully-Fisher method of Davis, Nusser, \& Willick
1996), and very small smoothing lengths (the VELMOD method of Willick
\etal 1997; Willick \& Strauss 1998).  Davis \etal (1991), and more
recently Berlind \etal (2000) discuss the systematic errors caused by
mismatch of smoothing scales between the velocity and density fields.
In this paper, we concentrate on the velocity--velocity comparison
after smoothing on scales of 4 \hmpc\ or larger: any smaller
would be affected by strongly nonlinear effects, while larger
smoothing would reduce the number of independent volumes over which
the comparison could be made.

The amplitude of the velocity field smoothed on a given scale $R$ with
the window $W$ depends on the density field power spectrum, $P(k)$, as
\be \langle v^2\rangle \propto \int dk P(k)
\widetilde{W}^2(kR)\,, 
\label{eq:rmsvel} 
\ee 
while for the density field the relation is as follows:
\be \langle \de^2\rangle \propto \int dk k^2 P(k) 
\widetilde{W}^2(kR)\, 
\label{eq:rmsdel} 
\ee 
(see the discussion in chapter 2 of Strauss \& Willick 1995). Here,
$\widetilde{W}$ is the Fourier transform of the smoothing window. The
absence of the $k^2$ term in equation~(\ref{eq:rmsvel}) means that the
velocity field is more heavily weighted by modes with low values of
the wavenumber $k$, i.e., large scales which are fully in the linear
regime. Therefore, we expect the relation between the velocity,
$\bfv$, and the gravity, $\bfg$, to be closer to linear than that
between the density and velocity divergence.


KCPR have shown that the relation between
\be
\te \equiv  -(H_0f)^{-1}\nabla \cdot \bfv(\bfr)
\,,
\label{eq:theta}
\ee
(see equation~\ref{eq:lindif}) and $\de$ (proportional to $\nabla
\cdot \bfg(\bfr)$) is nonlinear on small scales, and have proposed a
semi-empirical formula accurately describing the dependence of
$\theta_\delta \equiv \left\langle \theta | \delta\right\rangle$ on
$\de$:
\be
\theta_\delta = \alpha \left[ (1+\delta)^{1/\alpha} -1 \right]
+\epsilon
\,.
\label{eq:alf}
\ee
Here, the constant $\epsilon$ is introduced in order to
keep $\langle \theta \rangle = 0$ as it must; it is approximately
\be
\epsilon = \frac{\alpha-1}{2\alpha} \sigma_\de^2 + 
\calO\left(\sigma_\de^4\right)
\,,
\label{eq:eps}
\ee
where $\sigma_\de^2 \equiv\langle \delta^2\rangle$ denotes the
variance of the density field. Thus, this approximation to $\epsilon$
gets progressively worse as the fluctuations become more nonlinear and
in the following we use the exact value of $\epsilon$, obtained
numerically. KCPR have found that $\alpha = 1.9$ is a good fit over a
large range of smoothing scales.
Solving  Equation~\ref{eq:theta} for $\bfv$ in
the case of an irrotational flow gives:
\be
 \bfv(\bfr)=
   \frac{H_0 f}{4\pi}\int {\rm d}^3\bfr'
   \frac{\theta_\delta(\bfr')(\bfr'-\bfr)}
        {|\bfr'-\bfr|^3}
 \,,
 \label{eq:theint}
\ee
where equation~(\ref{eq:alf}) gives an expression for $\theta_\delta$.

In this paper, we investigate the nonlinearities of the relationship
between the velocity and gravity fields using a set of grid-based
simulations. The simulations are described in \S 2.  In
\S 3, we ask how well the probability distribution functions (PDFs) of
the Cartesian components of $\bfv$ and $\bfg$ are fitted with a
Gaussian.  Given that the source fields for the velocity and gravity
fields (i.e. the velocity divergence and density contrast
respectively) have mildly non-Gaussian distributions, it is not a
priori obvious what the distribution of the integral quantities should
be. In \S\ref{sec:vg} we directly measure the relation between $\bfv$
and $\bfg$ on various scales, and test the extent to which linear
theory, or non-linear extensions to it, may hold.  This is important
in determining whether the existing velocity--velocity comparisons
which use linear theory give biased results. We present our
conclusions in \S 5. Two appendices contain derivations of results
used in the text.

\section{The simulations}

We performed our simulations using the CPPA (Cosmological Pressureless
Parabolic Advection) code (see Kudlicki \etal 1996, KCPR). Matter in
this code is represented as a non-relativistic pressureless fluid, and
its equations of motion are solved on a uniform grid fixed in comoving
coordinates. Periodic boundary conditions are applied. Parabolic
interpolation of the hydrodynamical state (like in the Piecewise Parabolic
Method scheme, see Colella \& Woodward 1984) assures low internal
diffusion of the code and accurate treatment of high density and
velocity gradients.

We chose to use a grid-based code rather than an $N$-body code,
because it produces a volume-weighted velocity field directly. This is
important because equation (\ref{eq:theint}) is a solution to
equation~(\ref{eq:theta}) only when $\bfv$ is a potential (curl-free)
field, and the {\em mass-weighted} velocity field exhibits curl even
in the linear regime.\footnote{The mass-weighted velocity field is
proportional to $(1+\de)\bfv$.  If its curl is to be zero, with
$\nabla \times \bfv = 0$, one would require $\nabla\de\times\bfv =0$.
There is no a priori reason for this even in the linear regime.  The
mean value of the cosine of the angle between $\nabla\de$ and $\bfv$
measured on the 30 \hmpc\ scale in our six high-resolution
simulations (see Table 1) is $0.81$. The directions of these vectors are thus
correlated, but not parallel.} Moreover, the a priori unknown galaxy
bias does not allow one to treat observational data as purely
mass-weighted anyway.

All our simulations assume an Einstein-de Sitter universe. The
relation between the velocity and the (scaled) gravity in the mildly
nonlinear regime is insensitive to the Cosmological Density Parameter
and Cosmological Constant, as demonstrated both analytically (Gramann
1993, Chodorowski 1997, Nusser \& Colberg 1998; see also App.~B.3 of
Scoccimarro \etal 1998), and by means of N-body simulations
(Bernardeau \etal 1999), thus our results should be valid for any
cosmology.

%

Our simulations start from Gaussian density fluctuations with the
linear APM power spectrum (Baugh \& Efstathiou 1993, 1994, Baugh \&
Gazta\~naga 1996). The initial velocity field is obtained from
equation~(\ref{eq:linint}). The initial fields are evolved until the
linear variance of the density in spheres of radius 8 \hmpc,
$\s_8$, is unity. Then for subsequent analysis the output is selected
for which nonlinear $\s_8 = 0.87$ (the cluster normalization in
the currently preferred model $\Omega_m = 0.3$, $\Omega_\Lambda = 0.7$;
Eke, Cole \& Frenk 1996).

In order to test the dependence of the results on the spatial
resolution and box size we have studied three numerical models with
parameters given in Table \ref{tab:models}. To improve the statistics,
we have performed six realizations of each of the models, with
different random phases of the initial density field. We have also
performed a few additional simulations with the timestep reduced five
times; the results did not change noticeably. 

\begin{table}
\begin{center}
\caption{Parameters of the studied models. Hereafter, the model with
the cell size equal to $1.56$ \hmpc\ will be called the `high-resolution
model', while the remaining models will be called `low-resolution
models'.
\label{tab:models}}
\begin{tabular}{|c|c|c|c}
\hline
grid & box~size~[$h^{-1}\rm Mpc$] & cell~size~[$h^{-1}\rm Mpc$] \\
\hline \hline
$64^3$      &  200    &    3.13  \\

$128^3$     &  200    &    1.56  \\

$128^3$     &  400    &    3.13  \\
\hline
\end{tabular}
\end{center}
\end{table}

\section{The marginal distributions of v and g}
\label{sec:mar} 
The typical correlation length of the density field is of the order of
$5$ \hmpc, while $\bfg$ is influenced by density fluctuations in a much
larger region. For instance, the gravitational acceleration of the
Local Group receives considerable contributions from shells up to at
least 150 \hmpc\ (see e.g. Rowan-Robinson \etal 2000). Thus, $\bfg$, and
similarly $\bfv$, come from integration over an effective domain
containing a large number of essentially independent regions. Hence,
the Central Limit Theorem suggests that they should have close to
Gaussian distributions.

\begin{figure}
\vspace{-0mm}
\centerline{
\hspace*{15mm}
\hfill
\epsfxsize=100mm\epsffile{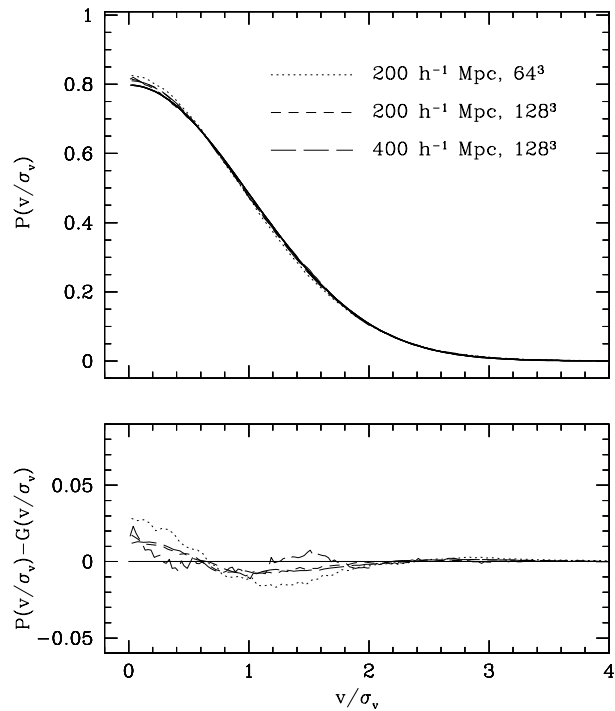}
\hfill
           }
\vspace{-0mm}
\caption{
The probability distribution of a Cartesian component of the 4 \hmpc\
Gaussian-smoothed velocity field from our simulations and its
deviations from Gaussianity. In both panels, the dotted line
represents the results for the $64^3$ model in the 200
\hmpc\ box, dashed --  the $128^3$ model in the 200 \hmpc\ box, 
and long-dashed -- the  $128^3$ model in the 400
\hmpc\ box. The solid line shows a standardized Gaussian, $G(v/\sigma_v)$.
\label{fig:pdfv5}}
\end{figure}

\begin{figure}
\vspace{-0mm}
\centerline{
\hspace*{15mm}
\hfill
\epsfxsize=100mm\epsffile{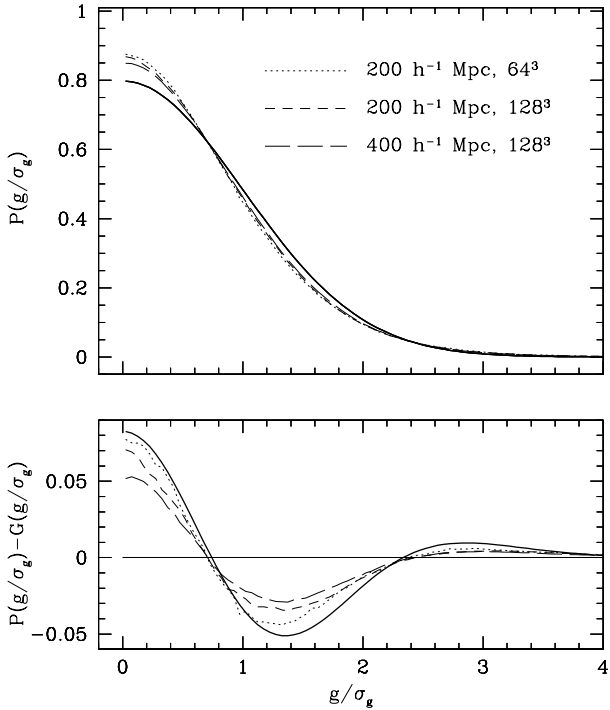}
\hfill
           }
\vspace{-0mm}
\caption{
The probability distribution of a Cartesian component of the 4 \hmpc\
Gaussian-smoothed gravity field from our simulations and its
deviations from Gaussianity. Line coding is as in
Figure~\ref{fig:pdfv5}. Thick solid line in the lower panel shows the
prediction of equation~(\ref{eq:Edge}), with $\calK_{g} = \calS_{4 g}
\s_\de^2 = 0.83$. 
\label{fig:pdfg5}}
\end{figure}

We test the Gaussianity hypothesis with our simulations.  We plot the
measured distribution functions for individual Cartesian components of
the peculiar velocity and gravity fields, which we label $v$ and $g$.
It turns out that on mildly nonlinear scales, the distributions are
well-fitted by Gaussians.  Figures~\ref{fig:pdfv5} and \ref{fig:pdfg5}
show the distributions for 4 \hmpc\ Gaussian smoothing. (As there is
no preferred direction in space, the distributions must be even
functions. Therefore, without the loss of information we plot them as
functions of the absolute value of a Cartesian component.) The
closeness to a Gaussian distribution is remarkable, especially for the
velocity field.

A standard way of quantifying modest departures from Gaussianity is to
decompose the PDF with an Edgeworth expansion: a leading Gaussian,
plus correction terms with amplitudes proportional to the higher-order
connected moments of the field (Longuet-Higgins 1963, Juszkiewicz
\etal 1995, Bernardeau \& Kofman 1995). As there is no preferred
direction in space, the skewness of the PDF of the Cartesian
coordinates of the velocity and gravity fields must equal zero. Thus
the first non-vanishing connected moment of a Cartesian component of
the velocity field is its kurtosis, $\calK_v = (\langle v^4 \rangle -
3\langle v^2 \rangle^2)/\langle v^2 \rangle^2$, and similarly for the
gravity field.  Therefore, the kurtosis measures the leading-order
departure from Gaussianity of the fields (Kofman \etal 1994, Catelan
\& Moscardini 1994). Specifically, the Edgeworth expansion in our case
reads:

\be
P(\mu) = \phi(\mu) \left[1 + \f{\calS_{4} \s_\de^2}{24} H_4(\mu) +
\calO\left(\s_\de^4 \right)\right] .
\label{eq:Edge}
\ee
Here, $P(\mu)$ is the PDF for the variable $v/\s_v$ or $g/\s_g$, where
$\s^2_v = \lan v^2 \ran$ and $\s^2_g = \lan g^2 \ran$ are the variances
of the velocity and gravity fields respectively, and $\phi(\mu) =
(2\pi)^{-1/2} \exp(-\mu^2/2)$ is the standardized normal
distribution. The symbol $H_4$ denotes the fourth order Hermite
polynomial. The quantity $\calS_4$ is related to the velocity, or
gravity, kurtosis and the variance of the {\em density} field in the
following way:

\be
\calK_{v} = \calS_{4 v} \s_\de^2 \,, \qquad \hbox{and} \qquad 
\calK_{g} = \calS_{4 g} \s_\de^2 \,.
\label{eq:S_4}
\ee
The quantities $\calS_{4 v}$ and $\calS_{4 g}$ are called hierarchical
amplitudes. For a given smoothing scale, during the weakly nonlinear
evolution they are constant, independent of the normalization of the
power spectrum.

\begin{figure*}
\vspace{-0mm}
\centerline{
\hspace*{0mm}
\hfill
\epsfxsize=80mm\epsffile{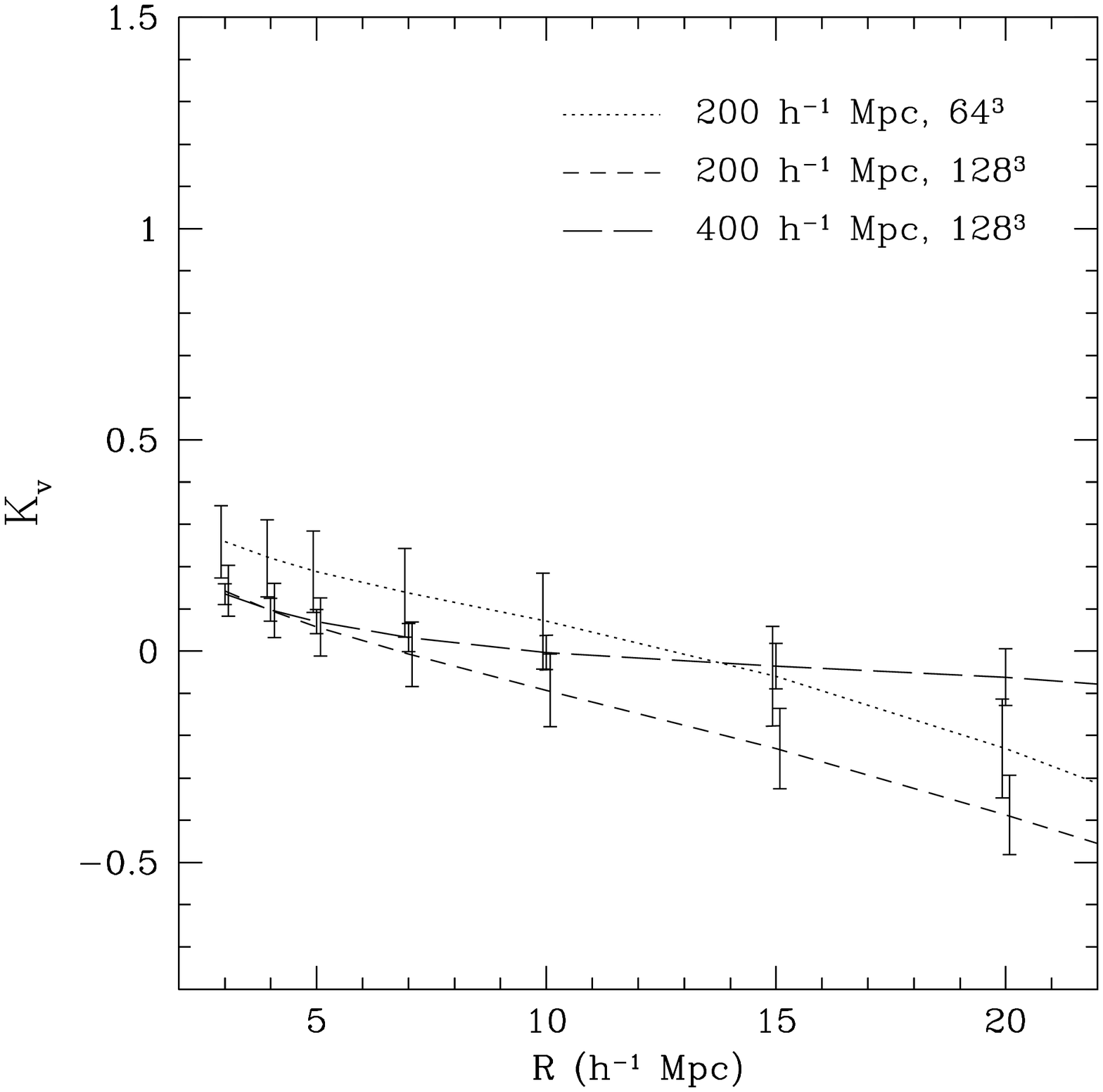}
\hfill
\epsfxsize=80mm\epsffile{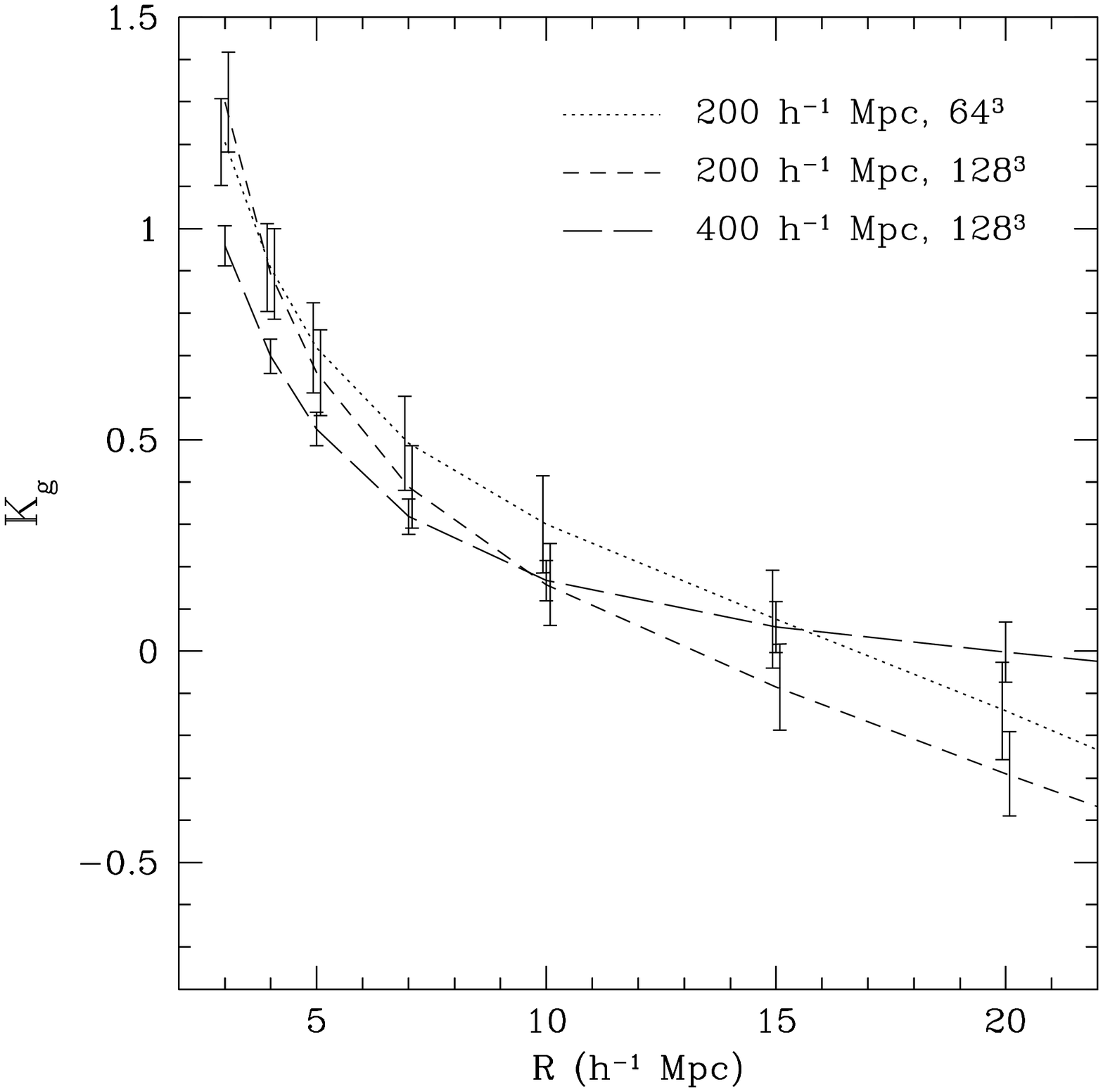}
\hfill
           }
\vspace{-0mm}
\caption{
Kurtosis of a Cartesian component of the velocity field (left) and gravity
field (right) from our simulations as a function of the Gaussian
smoothing length. The points are averages over six simulations times three
directions, while the error bars are the standard deviations over this
ensemble.
\label{fig:k4v}
}
\end{figure*}

We plot $\calK_v$ and $\calK_g$ as a function of the Gaussian
smoothing radius, $R$, in Figure \ref{fig:k4v}. We find that $\calK_v$
is close to zero for smoothing between 3 and 15 \hmpc. In contrast,
the gravity field develops a detectable kurtosis. Still, this is
substantially less kurtosis than the kurtosis of the gravity's source
field, density. Measured from the high-resolution simulations (i.e.,
with the grid size $128^3$ and the box size of 200 \hmpc), for the 4
\hmpc\ smoothing the density kurtosis is $\calK_\de = 22.2 \pm 5.0$,
thus more than an order of magnitude bigger than the gravity
kurtosis. (For the same smoothing, the kurtosis of the velocity
divergence is $\calK_\te = 4.1 \pm 0.6$). Therefore, qualitatively the
picture is clear: our simulations reveal substantially less kurtosis
of the velocity and gravity fields than those of the velocity
divergence and density contrast respectively; consequently the
distributions of $\bfv$ and $\bfg$ are much closer to Gaussian. Our
findings are inconsistent with the results of perturbative
calculations of Catelan \& Moscardini (1994), and consistent with the
results of N-body simulations of the velocity field of Kofman \etal
(1994).\footnote{The kurtosis of the velocity field, computed
perturbatively by Catelan \& Moscardini (1994), is greater than unity
already for a Gaussian smoothing scale as large as 10 \hmpc\ for all
cosmological models they have considered, and grows with decreasing
smoothing scale approximately like $\s_\de^2(R)$. In contrast, Kofman
\etal (1994) find that {\sl `the PDF of velocity (\ldots) is almost
indistinguishable from Gaussian in the simulations'}.}

Quantitatively, the agreement of the evolved velocity and gravity PDFs
with the perturbatively motivated equation~(\ref{eq:Edge}) is less
than perfect. In the lower panel of Figure~\ref{fig:pdfg5} we plot the
difference between the measured distribution for a Cartesian component
of the gravity field and the standardized Gaussian. As the thick solid
line we plot the difference predicted from equation~(\ref{eq:Edge}),
with the value $\calK_g = 0.83$ (the mean value for all models). The
predicted difference has somewhat too high amplitude.

\begin{figure}
\vspace{-0mm}
\centerline{
\hspace*{15mm}
\hfill
\epsfxsize=100mm\epsffile{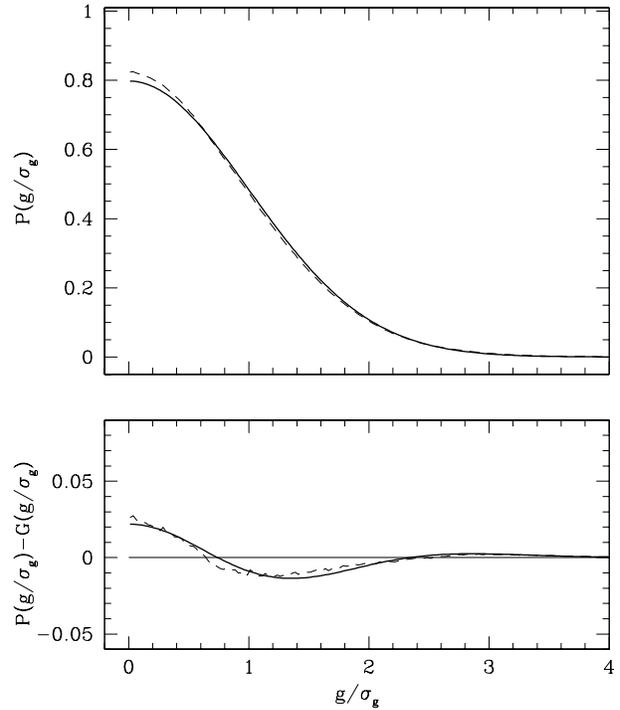}
\hfill
           }
\vspace{-0mm}
\caption{
The probability distribution of a Cartesian component of the 4 \hmpc\
Gaussian-smoothed gravity field from high-resolution simulations, for
$\s_\de^2 = 0.25$. In the upper panel, the dashed line shows the
simulated PDF while the thin solid line -- a standardized Gaussian. In
the lower panel, the thick solid line shows the prediction of
formula~(\ref{eq:Edge}), with $\calK_g = 0.22$. 
\label{fig:resid_025}}
\end{figure}

Expansion~(\ref{eq:Edge}) is an expansion in the density variance. In
our simulations, for 4 \hmpc\ smoothing $\s_\de^2 = 0.65$. For such a
big variance, higher-order contributions to formula~(\ref{eq:Edge})
may simply not be negligible. We have checked this conjecture by
analyzing the gravity field at an earlier output time. In
Figure~(\ref{fig:resid_025}), we plot the gravity PDF for $\s_\de^2 =
0.25$. Formula~(\ref{eq:Edge}) fits the simulated
distribution better.

We have also checked the perturbative scaling of the gravity (and
velocity) kurtosis with the variance of the density field, i.e.\
equation~(\ref{eq:S_4}). Although for a given smoothing scale, during
weakly non-linear evolution the hierarchical amplitude $\calS_{4g}$
remains constant, it depends moderately on the smoothing scale via an
effective spectral index of the power spectrum at this
scale. However, we have performed an additional simulation for a
power-law initial spectrum ($n=-1$), and found qualitatively similar
results; a change of sign in the kurtosis at several megaparsecs.   

\begin{figure}
\vspace{-0mm}
\centerline{
\hspace*{0mm}
\hfill
\epsfxsize=80mm\epsffile{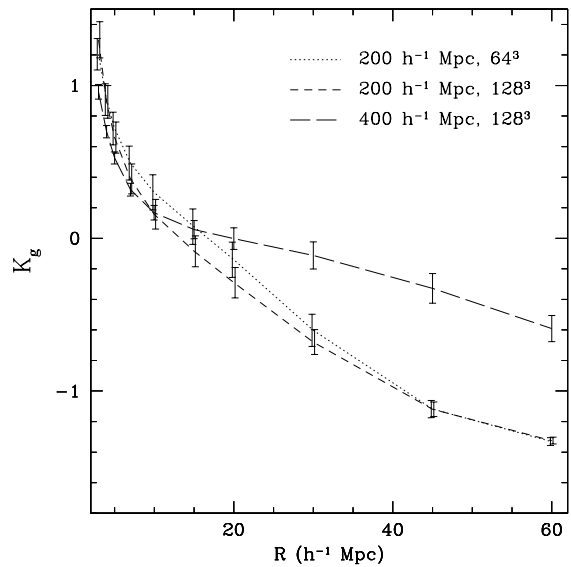}
\hfill
           }
\vspace{-0mm}
\caption{
Kurtosis of a Cartesian component of the gravity field from our
simulations as a function of the Gaussian smoothing length, now over a
larger range of smoothing scales. The points are averages over six
simulations times three directions, while the error bars are the
standard deviations over this ensemble. Line coding is as in
Figure~\ref{fig:pdfv5}. At large scales, the results for simulations
with the box size of 200 \hmpc\ converge to the same values,
regardless the resolution.
\label{fig:k4vall}}
\end{figure}

Intrigued by this feature, we have studied the values of the kurtoses
for even larger, apparently linear, scales. The results for the
gravity kurtosis are shown in Figure~\ref{fig:k4vall}. (The results
for the velocity kurtosis are similar.) The values of the kurtosis
clearly depend on the box size used in the simulations. For the box
size equal to 200 \hmpc, they tend, instead to zero, to an asymptotic
value $-1.5$. We interpret this as a purely numerical effect. Namely,
on large scales the velocity field of the simulation is dominated by a
small number of Fourier waves, and thus the Central Limit Theorem does
not force the distribution to be Gaussian, even in the initial
conditions. Equations (\ref{eq:rmsvel}) and (\ref{eq:rmsdel}) show
that this effect should be much stronger for $g$ than for $\de$. On
the largest scales, approaching the simulation box size, $\calK_g$
converges to $-1.5$, i.e.\ the value expected for a single mode, as
shown in Appendix \ref{app:a}.\footnote{As pointed out by Scherrer
(private communication), it is straightforward to show that the
kurtosis of a Cartesian component of an isotropic gravity field
satisfies $\calK_g \ge -6/5$, and similarly for the velocity
kurtosis. 
Therefore, any observations which violate this bound indicate that one
is in a regime in which the assumption of isotropy breaks down. This
happens in our simulations with the box size of 200 \hmpc\ for the
smoothing scale of $60$ \hmpc, at which $\calK_g < -1.2$ (see
Fig.~\ref{fig:k4vall}). This is consistent with our idea that for such
a big smoothing scale, a single mode dominates.} On scales shown
in Figure~\ref{fig:k4vall}, simulations with the 400 \hmpc\ box size
are much less affected by this effect, as expected.

Why is it that this 
seems to affect scales down to $\sim 20$ megaparsecs? Aren't there
enough modes with scales ranging from $\sim 20$ megaparsecs to the box
size, to warrant Gaussianity? The answer to this question is provided
by equation~(\ref{eq:rmsvel}). To the variance of the velocity field
on a given scale contribute all larger modes with amplitudes
proportional to $P(k)$, the density power spectrum. The spectrum of
the APM galaxies, employed by us, has a maximum at $\sim 300$
\hmpc. This is larger than the 200 \hmpc\ box size of some of our
simulations; thus there are relatively few modes on the scale of the
box size that dominate the velocity field on large scales.

To settle this issue definitely would require additional simulations
with box size comparable to the Hubble radius (like the Hubble Volume
simulations of the {\sc virgo} consortium, Evrard \etal
2002).\footnote{We have performed an additional simulation with the
grid size $256^3$ and the box size of 800 \hmpc. On scales shown in
Figure~\ref{fig:k4vall}, the gravity kurtosis tended asymptotically to
zero, remaining positive.} However, the ultimate goal of this paper is
to study the relation between velocity and gravity for 4 \hmpc\
smoothing.  Figure~\ref{fig:k4v} shows that on such a small scale the
velocity and gravity kurtoses are positive, thus most likely induced
by nonlinear dynamics rather than due to finite volume of our
simulations. However, given the limitations of the Edgeworth
expansion,\footnote{The Edgeworth expansion is an asymptotic
expansion, not guaranteed to converge. It is not even guaranteed to be
positive definite.} we would like to have an alternative measure of
non-Gaussianity. As such a measure we borrow the concept of {\em
negentropy} from information theory (e.g. Cover \& Thomas 1991,
Papoulis 1991).

We define the entropy $S[f]$ of a probability distribution $f$:
\be
S[f] = - \int f(x) \log f(x) dx\,.
\label{entr}
\ee
It can be shown that, for a given variance, the entropy is maximal for
Gaussian fields. Hence the difference between the entropy of a given
field and the entropy of a Gaussian of the same variance, the {\em
negentropy} $\calN[f]$, can be used as a measure of departure from
Gaussianity:
\be
\calN[f] = S[{\rm Gauss}] - S[f]\,.
\label{negentr}
\ee
The negentropy was calculated by numerically integrating the integral
of equation~(\ref{entr}), using a PDF binned over the range from
$-5\sigma$ to $5\sigma$.  We have found that this technique applied to
a Gaussian with the same range and binning yielded a negentropy of
less than $10^{-5}$, which assures us that our results are not
affected by numerical effects.

\begin{figure}
\vspace{-0mm}
\centerline{
\hspace*{0mm}
\hfill
\epsfxsize=80mm\epsffile{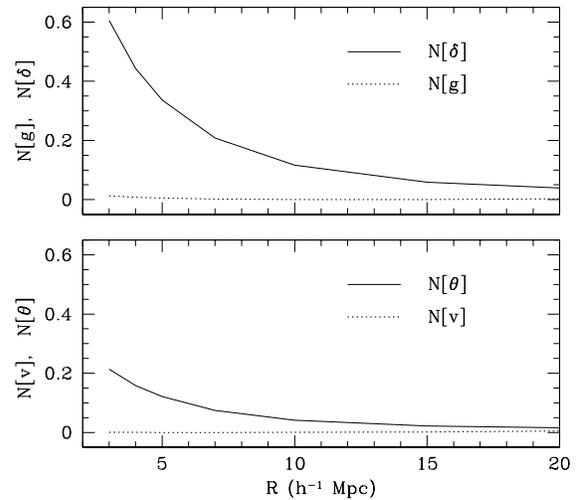}
\hfill
           }
\vspace{-9mm}
\caption{
Negentropies of the cosmic fields, for the high-resolution simulations
(i.e., with the grid $128^3$ and 200 \hmpc\ box size). Upper panel
compares the negentropies of gravity $g$ and of density contrast
$\de$, lower -- negentropies of velocity $v$ and of its divergence
$\te$.
\label{fig:neg}}
\end{figure}

We plot the results as a function of Gaussian smoothing scale in
Figure~\ref{fig:neg}. The negentropy of the density field is compared
to the negentropy of the gravity in the upper panel. The values for
velocity and its divergence are compared in the lower panel. The
negentropy of the velocity field is practically zero. The negentropy
of the gravity field, even on the smallest scales, is extremely small
compared to the negentropy of the density. The contrast between the
negentropy of the gravity and the negentropy of the density is even
greater than between the corresponding kurtoses. It is so because the
density field, unlike the gravity field, has significant skewness,
also contributing to the value of the negentropy. In other words, the
density field is more non-Gaussian than the gravity field not only
because it has bigger kurtosis, but also because it has non-vanishing
skewness. 

We conclude that on mildly non-linear scales, the non-Gaussianity of
$v$ and $g$ is completely negligible compared to the non-Gaussianity
of $\te$ and $\de$. As stated earlier, this finding is consistent with
the results of N-body simulations of the velocity field of Kofman
\etal (1994). It is also consistent with the results of Gooding \etal
(1992) and Scherrer (1992), where non-Gaussianity of the density field
was (at least on large scales) due to initial conditions rather than
nonlinear gravitational evolution.

\section{The one-component v--g relation}
\label{sec:vg}

\begin{figure}
\vspace{-0mm}
\centerline{
\hspace*{0mm}
\hfill
\epsfxsize=80mm\epsffile{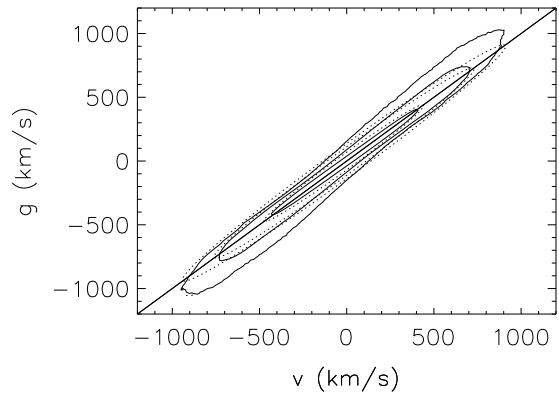}
\hfill
           }
\vspace{-0mm}
\caption{
The joint probability distribution of a Cartesian component of the
cosmic velocity and gravity fields, both smoothed with a 4 \hmpc\
Gaussian filter. Solid contours represent combined numerical results
from six high-resolution simulations, dotted ones -- a standardized
bivariate Gaussian of the same correlation coefficient. The contours
correspond to the probability levels of 68\%, 95\%, and 99\%.
\label{fig:gauss2d}}
\end{figure}


In grid simulations, the shortest Fourier modes correspond to the
Nyquist wavelength, of two cells. Smaller structures are not well
resolved by the code. In low-resolution simulations (cell size $3.13$
\hmpc, see Table~\ref{tab:models}), the Nyquist wavelength is larger
than the scale at which we would like to study the velocity--gravity
relation (4 \hmpc). In high-resolution simulations (cell size $1.56$
\hmpc), it is smaller.  Therefore, to model the velocity--gravity
relation at 4 \hmpc\ scale, we will use only high-resolution
simulations.

On scales of a few \hmpc, the relation between $\de$ and $\te$ is
non-linear. Nevertheless, since the probability distributions of the
Cartesian components of both $\bfv$ and $\bfg$ are nearly Gaussian, we
hypothesize that their joint probability distribution is a (bivariate)
Gaussian as well. In Figure \ref{fig:gauss2d} we present the simulated
joint PDF for $v$ and $g$, measured on the 4 \hmpc\ scale. It is
indeed quite close to a bivariate Gaussian. The only substantial
deviation is for the probability contour of 99\%, along the gravity
coordinate. There, positive kurtosis of the gravity field broadens the
simulated isocontour with respect to the Gaussian one. Figure
\ref{fig:gauss2d} also shows that $v$ and $g$ are strongly
correlated. In App.~\ref{app:gs} we show that in the bivariate
Gaussian case, the relationship between $v$ and $g$ is purely linear;
we now show that this is approximately the case in the simulations.

In the linear regime $\bfv=\bfg$ and thus each of the Cartesian
components of these quantities, which we denote $v$ and $g$,
respectively, are also equal.  In practice, the relationship between
these two quantities has some finite scatter (Fig.~\ref{fig:gauss2d}),
thus we will characterize the relation between the mean $v$ at
constant $g$, $\langle v | g \rangle$, and the converse, $\langle g |
v \rangle$.  Fitting a straight line to $\langle v | g \rangle$ as a
function of $g$ gives a slope of 0.94.  Therefore assuming pure
linear theory on this scale would give a systematic 6\% error in
$\beta$.  We now consider going beyond linear theory.  

The relationship between $\langle v | g \rangle$ and $g$
should be invariant to coordinate inversions, thus it must be an odd
function; similarly for the relationship between $\langle g | v
\rangle$ and $v$.

Hence, we shall adopt third-order polynomials
\be 
\langle g | v \rangle = c_1 v + c_3 v^3/\s_v^2
\label{eq:acr}
\ee
\be
\langle v | g \rangle = d_1 g + d_3 g^3/\s_g^2
\label{eq:adr}
\ee
 as the simplest odd non-linear model, and use the unitless parameters
$c_3$ and $d_3$ as a measure of the non-linearity of the relation. As
the deviations from linear theory are small, we expect $c_1$ and $d_1$
to be close to unity, and $c_3$ and $d_3$ to be small.  Since in
velocity--velocity comparisons one reconstructs velocities based on
the gravity field, the quantity $\langle v | g \rangle$ is more
relevant, and we shall concentrate on it in this paper. Note however
that relation (\ref{eq:acr}) can be used to transform velocities in
the first step of density--density comparisons such as POTENT (Dekel
\etal 1999).

\begin{figure}
\vspace{-0mm}
\centerline{
\hspace*{0mm}
\hfill
\epsfxsize=80mm\epsffile{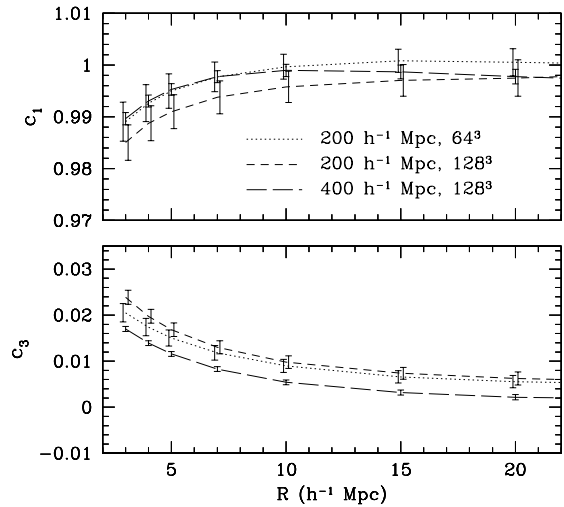}
\hfill
           }
\vspace{-9mm}
\caption{
The parameters $c_1$ and $c_3$ of the polynomial approximation to the
mean one-component $v$--$g$ relation (Eq. \ref{eq:acr}) as functions
of the smoothing scale. Line coding is as in Figure~\ref{fig:pdfv5}.
\label{fig:acr}}
\end{figure}

\begin{figure}
\vspace{-0mm}
\centerline{
\hspace*{0mm}
\hfill
\epsfxsize=80mm\epsffile{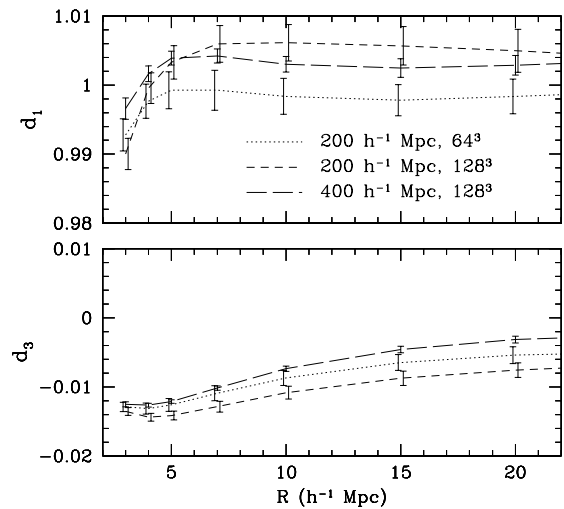}
\hfill
           }
\vspace{-9mm}
\caption{
The parameters $d_1$ and $d_3$ of the polynomial approximation to the
mean one-component $g$--$v$ relation (Eq. \ref{eq:adr}) as functions
of the smoothing scale. Line coding is as in Figure~\ref{fig:pdfv5}.
\label{fig:adr}}
\end{figure}

The parameters $c_1$ and $c_3$ as functions of the smoothing scale are
shown in Figure~\ref{fig:acr}, while Figure \ref{fig:adr} shows $d_1$
and $d_3$. We have fitted them independently for the three coordinates
in each of the six realizations of every numerical model; the lines
are the averages over these 18 datasets and errorbars indicate their
standard deviations. As expected, $c_3$ and $d_3$ are much smaller
than unity. On large scales, $c_1$ and $d_1$ tend to the linear theory
value with accuracy better than half per cent. Moreover, they remain
close to unity even on small scales, which implies that the systematic
bias in estimating $\beta$ based on the linear theory approximation
(\ref{eq:linint}) is small.

\begin{figure}
\vspace{-0mm}
\centerline{
\hspace*{0mm}
\hfill
\epsfxsize=80mm\epsffile{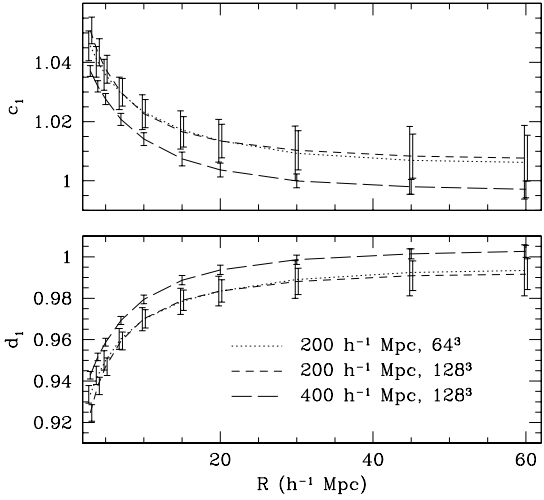}
\hfill
           }
\vspace{-9mm}
\caption{
The parameters $c_1$ and $d_1$ when the parameters $c_3$ and $d_3$ are
set to zero. Line coding is as in Figure~\ref{fig:pdfv5}.
\label{fig:c1d1}}
\end{figure}

To quantify this bias, in Figure~\ref{fig:c1d1} we show the parameters
$c_1$ and $d_1$ when $c_3$ and $d_3$ are set to zero. For $R > 20$
\hmpc, $c_1$ and $d_1$ deviate from unity by less than 2\%. Therefore,
at these scales one can apply linear theory with such good accuracy.

Note in Figures~\ref{fig:acr} and~\ref{fig:adr} that at small scales,
high-resolution simulations yield the highest values of $c_3$ and $1 -
c_1$ (similarly for $|d_3|$ and $1 - d_1$), but overall the effects of
resolution and of the simulation box size are not large. 

\begin{figure}
\vspace{-0mm}
\centerline{
\hspace*{10mm}
\hfill
\epsfxsize=109mm\epsffile{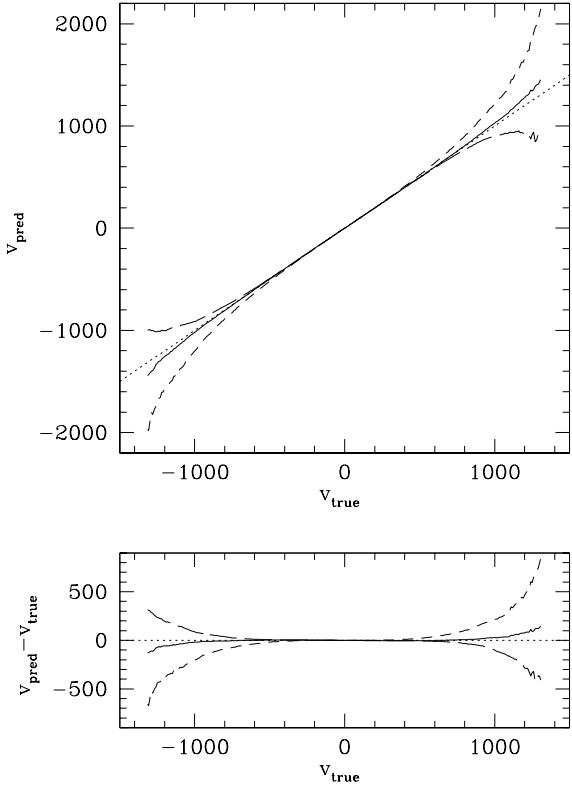}
\hfill
           }
\vspace{-0mm}
\caption{
One Cartesian component of the velocities reconstructed from the
density field -- combined data from six high-resolution simulations,
observed with a 4 \hmpc\ Gaussian filter. Top: mean predicted velocity
given true velocity. Bottom: errors of the reconstruction. All
velocities are in kilometers per second. The one-dimensional rms true
velocity, $\sigma_v$, for this smoothing equals 290 \kms. Dotted line:
identity, dashed line: linear theory model, long-dashed: a polynomial
model (Eq.~\ref{eq:adr}). Solid line shows the reconstruction from the
density field computed using the non-linear $\alpha\gamma$-formula
(Eq.~\ref{eq:alf_gam}), with $\alpha=1.56$ and $\gamma = 1.054$.
\label{fig:mean4}}
\end{figure}

We plot the mean predicted velocity as a function of the true velocity
for a 4 \hmpc\ Gaussian filter in Figure~\ref{fig:mean4}. The velocity
predicted with the linear-theory model (i.e., the gravity) is plotted
as a dashed line. Combined data from six realizations of our
high-resolution model (three components each) are binned with respect
to the predicted velocity and averaged in the bins. All velocities are
in kilometers per second (the one-dimensional rms true velocity
$\sigma_v = 290$ \kms\ on this scale). A long-dashed line shows the
velocity predicted by the polynomial formula (\ref{eq:adr}). We see
that linear theory, $v = g$, is a good fit for velocities up to about
$2\sigma_v$. The polynomial approximation works well up to about
$3\sigma_v$, but fails in the tails too. Therefore, we decided to seek
a better estimator of velocity from density than a simple function of
gravity.

\begin{figure}
\vspace{-0mm}
\centerline{
\hspace*{0mm}
\hfill
\epsfxsize=80mm\epsffile{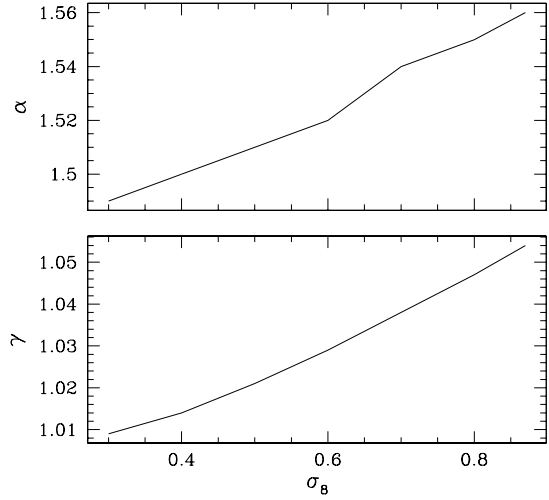}
\hfill
           }
\vspace{-9mm}
\caption{
The parameters $\alpha$ and $\gamma$ of formula~(\ref{eq:alf_gam}) for
4 \hmpc\ Gaussian smoothing, as functions of $\s_8$.
\label{fig:alf_gam}}
\end{figure}

KCPR have found that formula~(\ref{eq:alf}) is a good description of the tails
of the relation between the density and the velocity divergence, so it
should also work well on the integral level (eq.~\ref{eq:theint}). It
turned out, however, that at small scales a small modification is
needed. Formula~(\ref{eq:alf}) implies that the coefficient of the
linear term in density in the density--velocity divergence relation is
unity, $\te_\de = \de - r_2 (\de^2 - \s_\de^2) + \ldots$, where $r_2 =
(\alpha - 1)/2 \alpha$, regardless the value of $\alpha$, while we
have found here that at small scales it is not strictly true. To cure this
problem, we introduced an additional parameter, $\gamma$, so that
\be
\theta_\delta = \alpha \gamma \left[ (1+\delta)^{1/\alpha} -1 \right]
+ \epsilon\,.
\label{eq:alf_gam}
\ee
For given $\alpha$ and $\gamma$ the predicted velocity, $v_{\rm
pred}$, was calculated from equation~(\ref{eq:theint}). The best
values of $\alpha$ and $\gamma$ were found by minimizing the quantity
$X^2 \equiv \sum_i (v_{\rm pred}^{(i)} - v_{\rm true}^{(i)})^2 / (3
N^3)$, where $v_{\rm true}$ is the true velocity and the sum is over
all grid points ($N^3$ in total) and all Cartesian components. The
resulting values of the parameters, for the smoothing scale 4 \hmpc\
and different values of $\s_8$ at different output times, are shown in
Figure~\ref{fig:alf_gam}.

In the limit of linear theory, $\alpha = \gamma = 1$. The offset
$\epsilon$, fully determined by $\alpha$ and $\gamma$, is then equal
to zero. In general, on the integral level the value of $\epsilon$ is
irrelevant, since its contribution to velocity in the integral in
equation~(\ref{eq:theint}) averages out to zero. We see that for small
$\s_8$ the parameter $\gamma$ tends to unity, while $\alpha$ decreases
only weakly and remains well away from this value. This is consistent
with the findings of KCPR, that $\alpha$-formula (eq.~\ref{eq:alf}, or
eq.~\ref{eq:alf_gam} with $\gamma = 1$) describes well the
density--velocity relation in the weakly nonlinear regime (i.e., for
$\s_8$ smaller than, say, 0.3). Though the best-fit value of $\gamma$
is close to unity even for large $\s_8$, its inclusion markedly
decreases $X^2$.

\begin{figure}
\vspace{-0mm}
\centerline{
\hspace*{0mm}
\hfill
\epsfxsize=80mm\epsffile{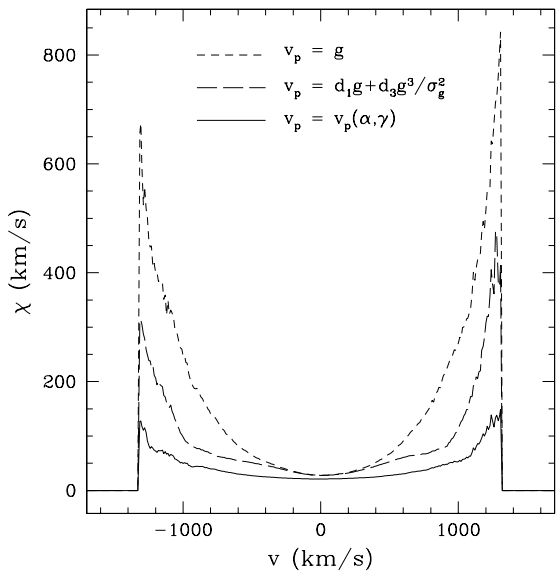}
\hfill
           }
\vspace{-0mm}
\caption{
The quantity $\chi$ (see text). Dashed line shows it for the velocity
predicted according to the linear theory, long-dashed -- for the
velocity approximated by a polynomial in the gravity, and solid -- for
the velocity predicted by the $\alpha\gamma$-formula (inserting
equation \ref{eq:alf_gam} in \ref{eq:theint}).
\label{fig:chi}}
\end{figure}

In Figure~\ref{fig:chi} we plot the quantity $\chi(v) \equiv
\sqrt{\Delta X^2/\Delta N}$, that is the square root of the contribution 
to $X^2$ from a given bin in velocity, divided by the number of points
in that bin. ($\chi$ has units of \kms.) We have $X^2 = \int
\chi^2(v)\, P(v)\, {\rm d}v$, where $P(v)$ is the probability distribution
function of a Cartesian component of the velocity field. Thus, $X^2$ is
a number-weighted average of $\chi^2$. Dashed line shows $\chi$ for
the velocity predicted according to the linear theory, long-dashed --
for the velocity approximated by a polynomial in the gravity, and
solid -- for the velocity predicted by the $\alpha\gamma$-formula
(inserting eq.~\ref{eq:alf_gam} in~\ref{eq:theint}). The linear-theory
estimator of velocity yields the largest values of $\chi$. The
polynomial formula results in smaller $\chi$ in the tails, but in
equal $\chi$ for `typical' values of velocity ($ -\s_v \le v \le
\s_v$). The $\alpha\gamma$-formula yields the smallest $\chi$ in the
whole range of velocity. For typical values of velocity, i.e.\ where
most of the data comes from, $\chi$ is slightly smaller than for the
previous estimators. Furthermore, it grows only moderately in the
tails.

In Figure~\ref{fig:mean4}, a solid line shows the mean velocity
predicted by the $\alpha\gamma$-formula as a function of the true
velocity. The smoothing scale is that of {\sc velmod}, i.e. 4 \hmpc,
and $\s_8 = 0.87$. For this value of $\s_8$, the best-fitted values of
$\alpha$ and $\gamma$ are $\alpha = 1.56$ and $\gamma = 1.054$. We see
that the $\alpha\gamma$-formula yields a practically unbiased estimator
of velocity. The difference between the predicted and true velocity in
the tails is very small. 

\begin{figure}
\vspace{-0mm}
\centerline{
\hspace*{0mm}
\hfill
\epsfxsize=80mm\epsffile{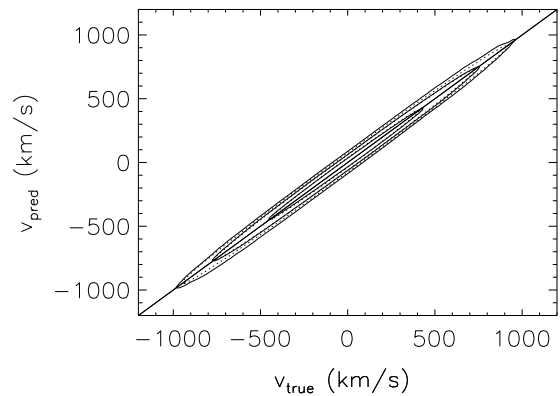}
\hfill
           }
\vspace{-0mm}
\caption{
The joint probability distribution of a Cartesian component of the
velocity predicted by the $\alpha \gamma$-formula and the true
velocity, both smoothed with a 4 \hmpc\ Gaussian filter. Solid
contours represent combined numerical results from six
high-resolution simulations, dotted ones -- a bivariate Gaussian of
the same correlation coefficient and variances. The contours
correspond to the probability levels of 68\%, 95\%, and 99\%.
\label{fig:vpred_vtrue}}
\end{figure}

Figure~\ref{fig:vpred_vtrue} demonstrates that the estimator of
velocity from density provided by the $\alpha\gamma$-formula is not
only essentially unbiased but also has a small variance (as suggested
already by Figure~\ref{fig:chi}). The figure shows a joint PDF for the
velocity predicted by the $\alpha\gamma$-formula and the true
velocity, for 4 \hmpc\ smoothing. It is instructive to compare it with
Figure~\ref{fig:gauss2d}, where $v_{\rm pred} = g$. We have found that
for larger values of the smoothing scale the correlation between the
predicted and true velocity is even higher. 

\begin{figure}
\vspace{-0mm}
\centerline{
\hspace*{0mm}
\hfill
\epsfxsize=80mm\epsffile{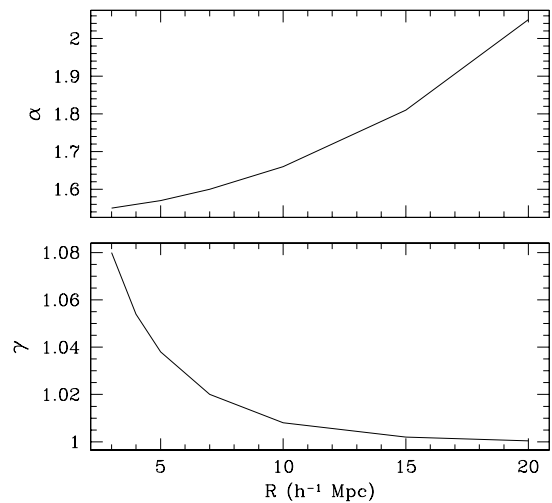}
\hfill
           }
\vspace{-9mm}
\caption{
The dependence of $\alpha$ and $\gamma$ on the smoothing scale, for
$\s_8 = 0.87$.
\label{fig:alfgam(R)}}
\end{figure}

In Figure~\ref{fig:alfgam(R)} we show the dependence of $\alpha$ and
$\gamma$ on the smoothing scale, for $\s_8 = 0.87$. For large $R$ the
parameter $\gamma$ tends to unity, as expected. The parameter $\alpha$
grows from the value around 1.55 for 3 \hmpc, to around 2.05 for 20
\hmpc. This is in contrast with the results of KCPR, where $\alpha
\simeq 1.9$ was a good fit for a large range of scales. We can see at
least three likely sources of this discrepancy. First, the power
spectrum used by KCPR was a pure power law, $P(k) \propto k^{-1}$,
while ours is not and the effective spectral index of our APM spectrum
depends (slightly) on the smoothing scale. Secondly, KCPR studied the
$\alpha$-formula, equation~{\ref{eq:alf}}, which is equivalent to our
$\alpha\gamma$-formula only when $\gamma = 1$. Finally, KCPR's fit was
obtained for all bins in velocity having equal weight, while we weight
the bins by the number of points. As a result, our fit is much less
affected by the tails of the velocity distribution.

To summarize, we have tested various estimators of velocity from
density. Nonlinear effects can be accounted for either at the integral
level, employing equation~(\ref{eq:adr}), or at the differential
level, employing equation~(\ref{eq:alf_gam}). We have shown that the
differential method works much better.


\section{Summary and conclusions}
\label{sec:end}

In this paper we have studied the cosmic velocity--gravity relation.
First, we have measured the non-Gaussianities of the cosmic velocity
and gravity fields, evolved from Gaussian initial conditions, by
computing their kurtoses and negentropies. We have shown that, on
scales of a few \hmpc, the non-Gaussianities of the cosmic velocity
and gravity fields are small compared to the non-Gaussianities of
velocity divergence and density. A similar finding for the velocity
field was reported by Kofman \etal (1994). Guided by this result, we
have shown that the relation between $v$ and $g$ is nearly
linear. Moreover, its proportionality coefficient is close to that
predicted by linear theory. Specifically, we have shown that the
systematic errors in velocity--velocity comparisons due to assuming
linear theory do not exceed 6\% in $\beta$. (Strictly speaking, if
$\s_8 < 0.9$ and the smoothing scale is not smaller than 4 \hmpc). To
correct for this small bias, we have tested various nonlinear
estimators of velocity from density. We have shown that the
$\alpha\gamma$-formula (a slight modification of the $\alpha$-formula
proposed by KCPR) yields an estimator which is essentially unbiased
and of small variance.

The smoothing of observed peculiar velocity data, with its sparse and
noisy coverage of the velocity, is technically difficult.  Thus
velocity--velocity analyses like {\sc velmod} compare unsmoothed
peculiar velocities with minimally smoothed predicted velocity fields
from redshift surveys.  The finite resolution of a grid code does not
allow us to test this effect; it would be worthwhile to repeat the
calculations with a high resolution $N$-body code, or with CPPA
enhanced with Adaptive Mesh Refinement.  The former has been done by
Berlind \etal (2000), but they did not separate the effects of
non-linear evolution from the effects of different smoothing of the
velocity and gravity fields.

An alternative method of estimating $\beta$ from cosmic velocities is
by comparing the cosmic microwave background dipole with the density
dipole inferred from a galaxy redshift survey. In maximum-likelihood
approaches to this problem (Strauss \etal 1992, Schmoldt \etal 1999,
Chodorowski \& Cieciel\c ag 2002) the relevant quantity is the joint
distribution for the velocity and gravity, which is commonly
approximated as a multivariate Gaussian. We have shown that this
indeed holds to fairly good accuracy (Fig.~\ref{fig:gauss2d}).  Small
nonlinear effects can be corrected for by a more thorough modelling of
the joint distribution. There is, however, a better approach: from a
redshift survey, instead of calculating the acceleration on the Local
Group (i.e., the density dipole, or gravity), one can calculate the
predicted velocity from the $\alpha\gamma$-formula. The joint
distribution for the predicted velocity and the true velocity is
perfectly Gaussian (see Fig.~\ref{fig:vpred_vtrue}). Moreover, the
relation between these two variables is tighter than the relation
between the velocity and the gravity. This implies that in the
proposed method, the estimated value of $\beta$ will be unbiased and
its inferred errors will be smaller.

\section*{Acknowledgments}
We thank Tomek Plewa for his contribution to the numerical code we use
and Micha\l\ R\'o\.zyczka for fruitful and animating discussions.
This research has been supported in part by the Polish State Committee
for Scientific Research grants No.~2.P03D.014.19 and 2.P03D.017.19,
and NSF grant AST96-16901. The simulations were performed at the 
{\em Interdisciplinary Centre for Mathematical and Computational
Modelling}, Pawi\'nskiego 5A, PL-02-106, Warsaw.

\appendix
\section{Calculation of the kurtosis of velocity for a single Fourier mode}
\label{app:a}

Let us assume a single--mode distribution of the density field.
All variables will depend on one dimension only.
In the linear regime the density field is
\be
\delta(x,y,z) = \delta(x)=\sin(x)\,. 
\label{eq:a1}
\ee
(The amplitude of the wave is irrelevant here, since
it will cancel out in the calculation of the dimensionless kurtosis.) 
Then, in the linear regime,
\be
\delta = \theta = - \nabla\cdot v = -\frac{{\rm d}v}{{\rm d}x} \,.
\label{eq:a2}
\ee
[$\theta$ denotes the scaled velocity divergence, so in
equation~(\ref{eq:a2}) there is no $f(\Omega_m,\Omega_\Lambda)$ term.]
Hence,
\be
v = -\int\delta(x)dx
\ee
and
\be
v(x) = \cos(x)\,.
\ee
Let $P(v)$ denote the volume-weighted probability distribution
function of a single component of the velocity field:
\be
P(v)|dv| = \frac{1}{\pi}|dx|\,.
\ee
Hence, 
\be
P(v)=\frac{1}{\pi\sin(x)}=\frac{1}{\pi\sqrt{1-\cos^2(x)}}\,,
\ee
and finally
\be
P(v)=\pi^{-1} \left(1-v^2\right)^{-1/2}\,.
\ee
Now let us compute the four first moments of this distribution:
\be
\langle v \rangle = 0
\ee
\be
\langle v^2 \rangle = \frac{1}{\pi}\int_{-1}^1\frac{v^2 dv}{\sqrt{1-v^2}} =
 \frac{1}{2}
\ee
\be
\langle v^3 \rangle = 0
\ee
\be
\langle v^4 \rangle = \frac{1}{\pi}\int_{-1}^1\frac{v^4 dv}{\sqrt{1-v^2}} =
 \frac{3}{8}\,.
\ee
We can calculate the kurtosis of this field now:
\be
K_v = \frac{\langle v^4 \rangle - 3 \langle v^2 \rangle^2}
           {\langle v^2 \rangle^2}
    = -\frac{3}{2}
\ee

\section{The mean relation between two Gaussian variables}
\label{app:gs}

Consider two correlated Gaussian variables, $y_1$ and $y_2$, with zero
means and variances $\sigma_1^{2} \equiv \lan y_1^2 \ran$ and
$\sigma_2^2 \equiv \lan y_2^2 \ran$, respectively. Their correlation coefficient is
\begin{equation}        \label{l1}
r_{} \equiv \frac{\lan y_1 y_2 \ran}{\sigma_1 \sigma_2}.
\end{equation}

Let us introduce normalized variables, $\mu \equiv y_1/\sigma_1$ and $\nu
\equiv y_2/\sigma_2$. We require that their joint
probability distribution function (PDF) is a bivariate Gaussian,
\begin{equation}        \label{l2}
    p(\mu, \nu) = \frac{1}{2 \pi \sqrt{1 - r_{}^2}}
    {\rm exp} \left[- \frac{(\mu^{2} - 2 r_{} \mu \nu + \nu^{2})}{2
    (1- r_{}^{2})}  \right].
\end{equation}
Were the fields uncorrelated ($r_{}=0$), $p(\mu, \nu)$ would be just
a product of two Gaussian distributions of $\mu$ and $\nu$.

Since the variables are assumed to be correlated but in general not
identical, the relation between them will have a scatter. The mean
relation between $\mu$ and $\nu$ is defined as mean $\mu$ given $\nu$,
$\lan\mu|\nu\ran$. By definition, the conditional PDF of $\mu$ given
$\nu$, $p(\mu |\nu)$, is $p(\mu,\nu)/p(\nu)$. Equation~(\ref{l2})
yields
\begin{equation}        \label{l3}
    p(\mu|\nu) = \frac{1}{\sqrt{2 \pi (1 - r_{}^{2})}}
    \ {\rm exp} \left[- \frac{(\mu - r_{} \nu)^{2}}{2 (1- r_{}^{2})}
    \right].
\end{equation}
We see that the conditional PDF is a Gaussian with modified mean and
variance. Specifically, 
\begin{equation}        \label{l4}
\lan\mu|\nu\ran = r_{} \nu \,,
\end{equation}
and
\begin{equation}        \label{l5}
\lan\mu^2|\nu\ran - \lan\mu|\nu\ran^2 = 1- r_{}^{2} \,.
\end{equation}

The relation between two correlated Gaussian variables is thus
linear. If the variables are uncorrelated there is no relation
whatsoever.\footnote{This is only the case for Gaussian variables. For
a counterexample in the opposite case, consider $y_1$ and $y_2 =
y_1^2$, where $y_1$ is Gaussian.} Therefore, the relation between two
Gaussian variables, if it exists at all, is always linear.

One may define an `inverse' relation to that specified in
equation~(\ref{l4}), i.e., $\lan\nu|\mu\ran$. From symmetry of the
joint PDF,
\begin{equation}        \label{l6}
\lan\nu|\mu\ran = r_{} \mu \,.
\end{equation}
The proportionality coefficient in the `inverse' relation is thus not
a simple reciprocal of the coefficient in the `forward' relation,
equation~(\ref{l4}). The difference is equal to $(1-r^2)/r \simeq
1-r^2$ for $r$ close to unity. In other words, the difference between
the true and the `naive' coefficient of the inverse relation is
directly related to the scatter in the relation.

\end{document}